\documentclass{elsart}
\usepackage {graphicx}
\usepackage{amssymb}

\journal{Physica D: Nonlinear Phenomena}
\begin{document}

\begin{frontmatter}

\title{Dynamics of inhomogeneous one--dimensional coupled map lattices}

\author[1,2]{S.Rybalko} \ead{rybalko@chem.scphys.kyoto-u.ac.jp} and \author[1]{A.Loskutov}
\ead{loskutov@moldyn.phys.msu.ru}

\address[1]{Physics Faculty, Moscow State University, Moscow 119899 Russia}

\address[2]{Department of Physics, Graduate School of Science, Kyoto University,
Kyoto 606-8502, Japan}

\begin{abstract}

We study the dynamics of one--dimensional discrete models of
one--component active medium built up of spatially inhomogeneous
chains of diffusively coupled piecewise linear maps. The
nonhomogeneities (``defects'') are treated in terms of parameter
difference in the corresponding maps. Two types of space defects
are considered: periodic and localized. We elaborate analytic
approach to obtain the regions with the qualitatively different
behaviour in the parametric space. For the model with
space--periodic nonhomogeneities we found an exact boundary
separating the regions of regular and chaotic dynamics. For the
chain with a unique (localized) defect the numerical estimate is
given. The effect of the nonhomegeneity on the global dynamics of
the system is analyzed.

\end{abstract}

\end{frontmatter}

\section{Introduction}

One of the effective methods of analysis of several systems is based
on their representation by discrete models. For this one can use
the space discretization as well as the time one. Using the space
discretization the initial (continuous) system is modelled by a finite
set of elements that  are coupled according to some rules, which
follows from the nature of the system of interest. Each element of
these then occurs to be some (small) dynamical system which may be
described by a (small) set of dynamical variables. If the time is
also discretized, i.e. if the evolution of these dynamical
subsystems is considered at discrete time--steps, then such
subsystems are treated as maps. In this case spatio-tempotal
models are said to be {\it lattice} of coupled maps.

The examples of application of the space--time discretization in
studies of the nonlinear media are numerous. Here we mention the
problems related to chaos and noise in distributed systems
\cite{Kaneko89_1,Kaneko89_2,Busi93,Milos}, to description of the
excitable media \cite{Milos,Bare} and some other similar problems
(see e.g. \cite{Zhiweg,Morita,Munetc,Bunim,TheoryCML,ChaosCML,Kaneko90,Busi88,Buven}
and references therein). Moreover,
any  numerical analysis of a continuous system has in fact to be
reduced to the study of space--time discretized model.

In principle, a variety of different space--time lattices may be
used. The mostly studied are the models where only the
nearest--neighbour (local) coupling is taken into account
(see e.g. \cite{Munetc,Bunim,TheoryCML,Kaneko90,Busi88} and references
therein). In other models the global coupling, i.e. when all elements of
the system are coupled with each other, is assumed \cite{Pescer,Just,Kaneko91}.
In lattices with the local coupling, as a rule, this
is considered as a diffusional one \cite{Kaneko89_1,Kaneko89_2,Milos,Munetc,Bunim,TheoryCML,ChaosCML,Kaneko90,Busi88}.
Mainly, these models are used to analyse the spatio--temporal chaos,
pattern formation, self--organization phenomena, etc.
\cite{Kait,Bambiz,Mits}

If all the elements are the same, then such lattices are called
homogeneous. If, however, elements are different, then the lattice
is not a homogeneous one, and its investigation is much more
complex. With applied point of view the nonhomegeneity may
be two--fold: either elements of the system have different maps,
or they have functionally identical maps but with different control
parameters. The nonhomogeneities are usually said to be {\it defects}.
The relative position of defects may vary from a single localized
inclusion to a space--periodic ordering.

In most of studies of coupled map lattices, the case of homogeneous
lattices (and mainly numerically) has been addressed. However, it is
clear, that the assumption of the homogeneity is a crude
simplification. Therefore it is rather important to investigate the
sensitivity of the lattice dynamics to certain nonhomogeneities
(see, e.g., \cite{Shargup,Jopima} and references therein).

The present study is devoted to the spatially
inhomogeneous 1D lattices (in fact, chains) of diffusively coupled
piecewise--linear maps, which refer to some models of the
one--dimensional excitable media \cite{Afrneosh}. The defects are
determined as maps with different parameter values. Using the
exact calculation of the Lyapunov exponents we analyse the
behaviour of the lattice with a localized defect and with the
space--periodic nonhomegeneity. We also describe the structure of
the phase space of such systems.

\section{Models of spatially inhomogeneous excitable media}

Here we consider the dynamics of two models of inhomogeneous
active media: one--dimensional (annular) chain of coupled maps
with periodic nonhomegeneity and with one localized defect.
To introduce the notations, we remind that the term ``map'' is
implied transformation of some interval $I$ on itself: $T_a: I
\longrightarrow I$, where $I=[\alpha, \beta ]\in {\bf R}^1$,
$x\mapsto G(x,a)$, $x\in I$, $G(x,a)$ is some function and $a$ is
a parameter. In terms of iterations $k$ this may be written as
$x(k+1)=G(x(k),a)$, $x\in I$. The nonhomegeneity means
nonuniformity of the parameter $a$ along the chain of maps. The
functions $G$ are chosen in such a way that they model the
one--component active medium.

\subsection{Homogeneous medium}

\begin{figure}[ht]
\centering
\includegraphics[scale=0.8] {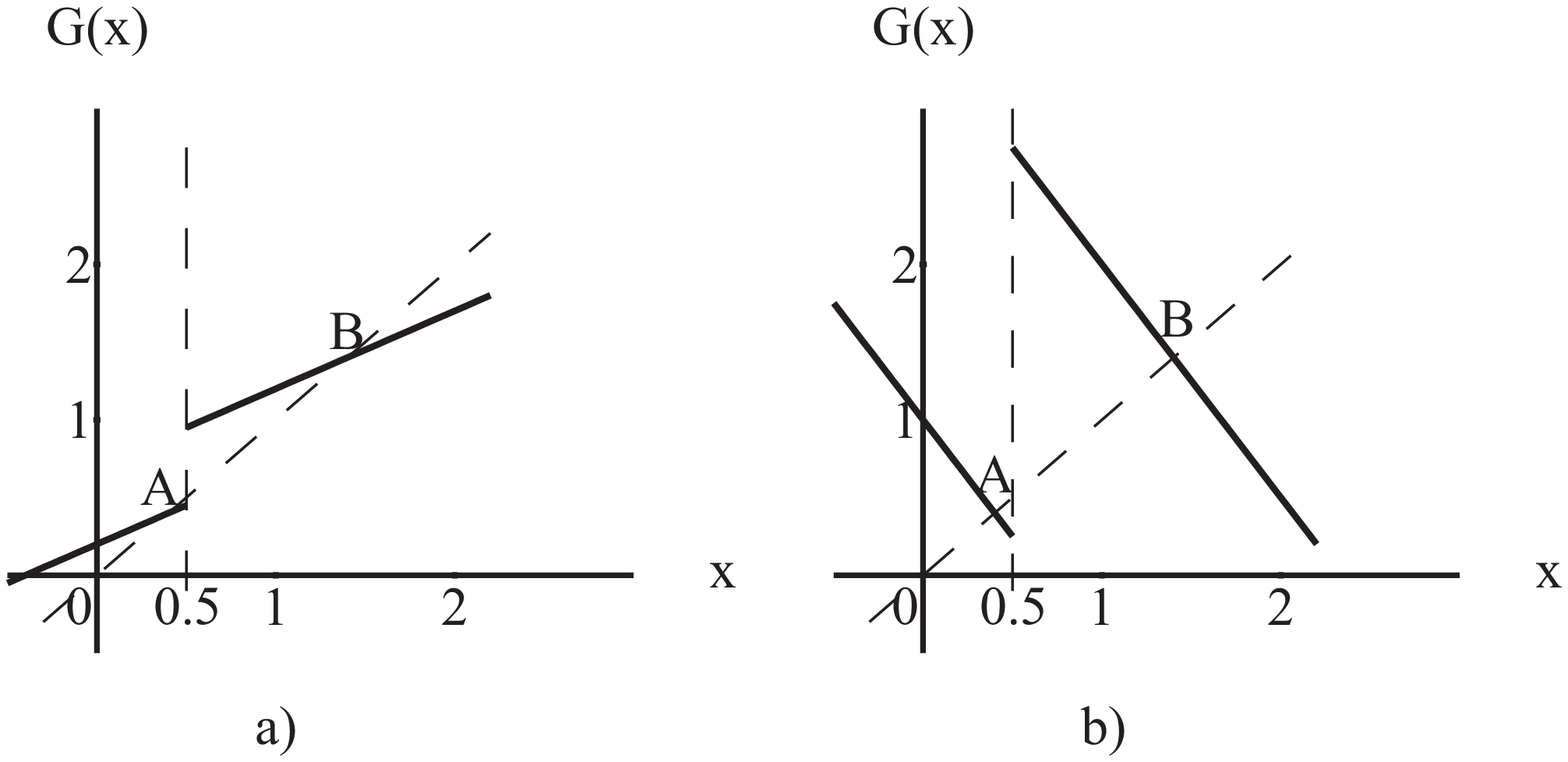}
\caption{The map $x\mapsto G(x,\alpha,\gamma)$ for
$\alpha=0.25$, $\gamma=-0.2$ (a) and $\alpha=1.25$,
$\gamma=-1$ (b).}
\label{mapg}
\end{figure}

Consider the system of $N$ diffusively coupled maps
\begin{equation}
\label{main}
x_n(k+1)=G(x_n(k), \alpha, \gamma)+
\varepsilon\Bigl(x_{n+1}(k)-2x_n(k)+x_{n-1}(k)\Bigr)
\end{equation}
with periodic boundary conditions, $x_{n+N}(k){\equiv }x_n(k)$,
where $n=1,2,\ldots,N$ being the discrete space coordinate,
$k=0,1,2,\ldots$ is the discrete time, $\varepsilon>0$ is the
(constant) diffusion coefficient, and $\alpha>0$, $\gamma$
are the parameters of the map. This choice of the coupling
means that the spatial and temporal actions in map (\ref{main}) occur
simultaneously. Let $G$ be the piecewise linear function:
\begin{equation}
\label{g}
G(x, \alpha, \gamma)=x+\alpha F(x)-\gamma=
\cases{
(1-2\alpha)x-\gamma, &$x<1/2$;\cr
\ &$\ $\cr
(1-2\alpha)x-\gamma+2\alpha, &$x>1/2$.\cr
}
\end{equation}
This function $G(x, \alpha, \gamma)$ is chosen in this form in order
to represent (in the discretized version) the basic model of the
one--component excitable medium, which is described by the
Kolmogorov--Petrovsky--Piskunov type equation:
$$
\frac{\partial x(z,t)}{\partial t}=D\frac{\partial^2x(z,t)}{\partial z^2}
+\Phi(x)\ .
$$
This equation is frequently used in biophysical problems, in the
combustion theory, in chemical kinetics (e.g. in the theory of the
Belousov--Zhabotinsky reaction), solid-state physics, etc. (see,
for example, \cite{Vasromyah,Autowdif,Zelbali,Arwein}).
One can see that the piecewise linear function
$G(x, \alpha ,\gamma )$ has a constant slope $(1-2\alpha )$.
Depending on $\alpha $ the map $x\mapsto G(x,\alpha, \gamma)$
may exhibit qualitatively different behaviour. For  $0<\alpha<1$
(see Fig.\ref{mapg}a) it gives rise to a regular dynamics with one
or two stable point, depending on the value of the parameter
$\gamma$ (points A and B in Fig.\ref{mapg}a), which are attractive
for almost all points of the phase space. For $\alpha >1$ the slope of
$G(x, \alpha, \gamma )$ is larger then one (see Fig.\ref{mapg}b).
That means, that for the finite motion of phase points the dynamics
would be chaotic.

The uniform model (\ref{main}) based on the coupled maps, which
are described by functions (\ref{g}), is known (see
\cite{Afrneosh}). Following \cite{Afrneosh} consider this system
as an $N$--dimensional piecewise--linear map $f: {\bf R}^N \to
{\bf R}^N$. Because $G(x, \alpha, \gamma)$ is the
piecewise--linear function with the constant derivative, the
differential  ${\rm D}_f$ of the map $f$ is a matrix with constant
coefficients. This easily allows us to find the Lyapunov exponents
$\lambda_i$ of such map. If $\rho_s$ are the eigenvalues of ${\rm
D}_f$, then  $\lambda_s=\ln|\rho_s|$. If among all the values
$\rho_s$ there exists an eigenvalue located outside the unit
circle on the complex plane, then any trajectory of the map $f$ is
unstable. Otherwise the model has the regular dynamics. Thus, to
know qualitatively the behavior of the system it is sufficient to
find the location of the eigenvalues $\rho_s$ on the complex
plane.

\begin{figure}[ht]
\centering
\includegraphics[scale=0.6] {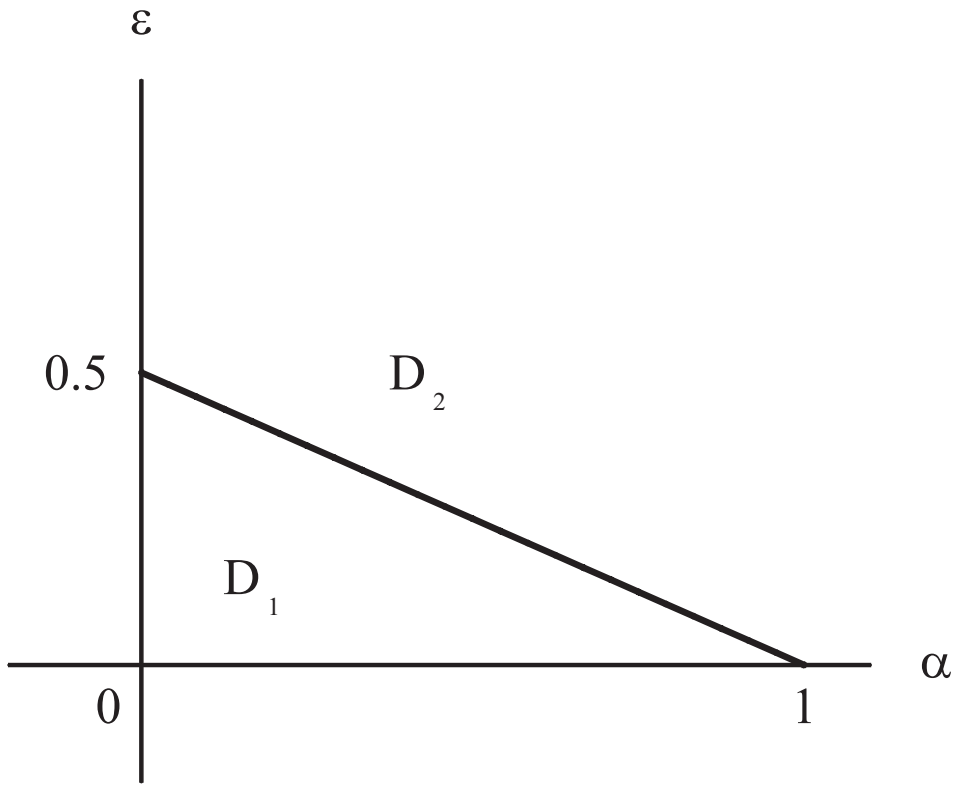}
\caption{The parametric space of the homogeneous annular
chain of the diffusively coupled maps (\ref{main}), (\ref{g}).}
\label{D1D2hom}
\end{figure}

To find $\rho_s$ we should analyse the characteristic polynomial
for the differential of the $N$--dimensional map, which may be
expressed via the determinants of three--diagonal matrices of
different dimension. For these determinants recurrent relations
may be obtained. Taking the recurrent relations as difference
equations with the given initial conditions we can find the
solutions which are proportional to the Chebyshev polynomials with
a linear function of the eigenvalues of ${\rm D}_f$
\cite{Afrneosh}. Using then the properties of the Chebyshev
polynomials it can be shown that the characteristic polynomial may
be reduced to the quadratic trinomial of a certain function
$\rho_s$. Calculation of $\rho_s$ and analysis of their location
on the complex plane shows that the parametric space of the system
(\ref{main}), (\ref{g}) has two regions with qualitatively
different behavior (Fig.\ref{D1D2hom}).

\begin{enumerate}

\item
The region $D_1$, with the values of the parameters $\alpha$ and
$\varepsilon$, which correspond to the regular dynamics of the model.
This implies that the modules of all the eigenvalues of the
differential of the map $f$ are less than unity.

\item
The region $D_2$ where at least one of the eigenvalues $\rho_s$
satisfies the condition $|\rho_s|>1$. Generally, the system may
have an infinite dynamics. However, if the dynamics is finite the
system (\ref{main}), (\ref{g}) behaves chaotically
\cite{Bunim,TheoryCML,ChaosCML}.

\end{enumerate}

The system (\ref{main}), (\ref{g}) is an uniform since all its
elements are identical (one has the same map $x\mapsto G(x,
\alpha, \gamma)$ for all the elements). Futher we
will assume that each element of the system has the same map
$x_i\mapsto G(x_i, \alpha_i, \gamma)$ but along the chain the
values of the control parameter $\alpha_i$ may differ\footnote{In
general, we can consider the case of different $\gamma$. But we
will see that the dynamics of both uniform and nonuniform chains
does not depends on $\gamma$.}. We assume that only two
types of elements exist in the system. The corresponding values we
denote as $\alpha$ and $\beta$. As for the case of uniform system
we will use the periodic boundary conditions, $x_{n+N}(k){\equiv
}x_n(k)$.

\subsection{The model with spatially periodic nonhomegeneity}

The periodic spatial nonhomegeneity implies in our case that the
chain has elements with alternating values of the parameters in
the map $G$, e.g. $\alpha, \beta, \alpha, \beta, \ldots\ $.
This particular nonuniform model allows a complete analytical
analysis.

Consider the system of coupled elements:
\begin{equation}
\label{period}
\cases{
x_n(k+1)=G(x_n(k), \alpha, \gamma)+
\varepsilon\Bigl(x_{n+1}(k)-2x_n(k)+x_{n-1}(k)\Bigr),
&$n$ is odd,\cr
\ &$\ $\cr
x_n(k+1)=G(x_n(k), \beta, \gamma)+
\varepsilon\Bigl(x_{n+1}(k)-2x_n(k)+x_{n-1}(k)\Bigr), &$n$
is even,\cr
}
\end{equation}
where $x_{n+N}(k)\equiv x_n(k)$, $N$ is even (to satisfy the
periodic boundary condition), and the function $G$ is given by
(\ref{g}). To analyse the dynamics of this system we calculate
the Lyapunov exponents using the approach described above.

The differential of the $N$--dimensional map corresponding to the
system (\ref{period}) reads
$$
{\rm D}_f \equiv Q_N =\left(
\begin{array}{ccccc}
1-2\varepsilon-2\alpha & \varepsilon & 0 &\ldots & \varepsilon\\
\varepsilon & 1-2\varepsilon-2\beta & \varepsilon &\ldots & 0\\
0 & \varepsilon & 1-2\varepsilon -2\alpha &\ldots & 0\\
\vdots & \vdots & \vdots & \ddots & \vdots\\
\varepsilon & 0 & 0 & \ldots & 1-2\varepsilon -2\beta
\end{array}
\right)\ .
$$
To find the eigenvalues $\rho_s$, $s=1,2,\ldots,N$ of the matrix
$Q_N$, we calculate the determinant
$$
{\rm det}(Q_N-\rho I_N)=\left|
\begin{array}{cccccc}
2z_1\varepsilon & \varepsilon & 0 & 0 & \ldots & \varepsilon\\
\varepsilon & 2z_2\varepsilon & \varepsilon & 0 & \ldots & 0\\
0 & \varepsilon & 2z_1\varepsilon & \varepsilon & \ldots & 0\\
\vdots & \vdots & \vdots & \vdots & \ddots & \vdots\\
0 & 0 & 0 & 0 & \ldots & \varepsilon\\
\varepsilon & 0 & 0 & 0 & \ldots & 2z_2\varepsilon
\end{array}
\right|\ ,
$$
with $z_1$ and  $z_2$ defined by the relations $1-2\varepsilon
-2\alpha -\rho =2z_1\varepsilon$, $ 1-2\varepsilon -2\beta -\rho
=2z_2\varepsilon$. Making expansion of the determinant
we obtain:
$$
{\rm det}(Q_N-\rho I_N)=2z_1\varepsilon
B_{N-1}-\varepsilon ^2B_{N-2}- 2\varepsilon ^N-
\varepsilon ^2\hat B_{N-2},
$$
where
$$
B_N=\left|
\begin{array}{ccccc}
2z_1\varepsilon & \varepsilon & 0 & \ldots & 0\\
\varepsilon & 2z_2\varepsilon & \varepsilon & \ldots & 0\\
0 & \varepsilon & 2z_1\varepsilon & \ldots & 0\\
\vdots & \vdots & \vdots & \ddots & \vdots\\
0 & 0 & 0 & \ldots & 2z_2\varepsilon
\end{array}
\right|\ , \ \ \ \
\hat B_N=\left|
\begin{array}{ccccc}
2z_2\varepsilon & \varepsilon & 0 & \ldots & 0\\
\varepsilon & 2z_1\varepsilon & \varepsilon & \ldots & 0\\
0 & \varepsilon & 2z_2\varepsilon & \ldots & 0\\
\vdots & \vdots &\vdots & \ddots & \vdots\\
0 & 0 & 0 & \ldots & 2z_1\varepsilon
\end{array}
\right|\ .
$$
It is not hard to see that $\hat B_N(z_1,z_2)$ may be obtained from
$B_N(z_1,z_2)$ exchanging the arguments $z_1\leftrightarrow z_2$.
This suggest the way of finding $B_N(z_1,z_2)$ using the recurrent
relation. Expanding $B_N$ (along the first line) we obtain:
$$
\cases{
B_N=2z_1\varepsilon B_{N-1}-\varepsilon ^2B_{N-2},
&$N$ is even;\cr
\ &$\ $\cr
B_N=2z_2\varepsilon B_{N-1}-\varepsilon ^2B_{N-2},
&$N$ is odd.
}
$$
This allows us to express determinants of the odd order via
determinants of the even one,
\begin{equation}
\label{four}
B_{2N+1}=\frac{B_{2N+2}+\varepsilon ^2B_{2N}}{2z_1\varepsilon }\ ,
\end{equation}
and obtain then the recurrent relation for the even determinants
\begin{equation}
\label{five}
B_{2N}=\varepsilon ^2(4z_1z_2-2)B_{2N-2}-\varepsilon ^4B_{2N-4}\ .
\end{equation}
One can treat Eq.(\ref{five}) as a difference equation with the initial
conditions
\begin{equation}
\label{six}
B_0=1, B_2=\varepsilon ^2(4z_1z_2-1)\ .
\end{equation}
Solving then (\ref{five}), (\ref{six}) together with (\ref{four}),
we arrive at:
$$
\cases{
B_N(z_1,z_2)=\varepsilon ^NU_N(z), &for even $N$;\cr
\ &$\ $\cr
B_N(z_1,z_2)=
\varepsilon ^N\sqrt{\displaystyle \frac{z_2}{z_1}}U_N(z),
&for odd $N$,\cr
}
$$
where $U_N(z)$ is the second order Chebyshev polynomial. In
turn, for $\hat B_N(z_1,z_2)$ we get:
$$
\cases{
\hat B_N(z_1, z_2)=\varepsilon ^N U_N(z), & for even $N$ ;\cr
\ &$\ $\cr
\hat B_N(z_1,z_2)=\varepsilon ^N
\sqrt{\displaystyle\frac{z_1}{z_2}}\ U_N(z),
&for odd $N$.\cr
}
$$
Now, using the obtained relations for $B_{N-1}$, $B_{N-2}$ and
$\hat B_{N-2}$, we finally arrive at:
\begin{equation}
\label{seven}
{\rm det}(Q_N-\rho I_N)=
2\varepsilon ^N\Bigl(zU_{N-1}(z)-U_{N-2}(z)-1\Bigr).
\end{equation}

This allows us to find the eigenvalues of the differential of the map
(\ref{period}), (\ref{g}). As follows from Eq.(\ref{seven}),
\begin{equation}
\label{eight}
zU_{N-1}(z)-U_{N-2}(z)-1=0,
\end{equation}
or, $T_N(z)-1=0$. Here we used the relation between the second
order Chebyshev polynomials $U_N(z)$ and the first order $T_N(z)$.
Since
$$ T_N(z)=\frac{\left(z-\sqrt{z^2-1}\right)^N+
\left(z-\sqrt{z^2-1}\right)^{-N}}{2},
$$
where $z^2=z_1z_2$, then
with $t=z-\sqrt{z^2-1}$ we obtain from
Eq.(\ref{eight}):
$$
\displaystyle\frac{1}{2}\left(t^N+\frac{1}{t^N}\right)-1=0
$$
or
$\displaystyle\left( t^N-1\right)^2=0$. Therefore,
$t_s=e^{i\frac{2\pi }{N}s}$, $s=1,2,\ldots ,N$, and $z_s=\cos(2\pi
s/N), s=1,2,\ldots ,N$. Taking into account that $2z_1\varepsilon
\cdot 2z_2\varepsilon =4z^2\varepsilon ^2$, we obtain a simple
equation: $$ (1-2\varepsilon -2\alpha -\rho )(1-2\varepsilon
-2\beta -\rho )= 4\varepsilon ^2\cos ^2\left(\frac{2\pi
}{N}s\right). $$ Solving it with respect to $\rho $, we find the
set of eigenvalues of the differential of the map (\ref{period}),
(\ref{g}):
\begin{equation}
\label{nine}
\rho ^s_{1,2}=1-2\varepsilon -\alpha -\beta \pm \sqrt{(\alpha -\beta )^2
+4\varepsilon ^2\cos ^2\left(\frac{2\pi }{N}s\right)}\ ,
\end{equation}
where $s=1,2,\ldots,N/2$. The dynamics of the system (\ref{period}),
(\ref{g}) would be completely regular if all the values of $\rho_s$ were
within the unit circle on the complex plane. Since all the
$\rho^s_{1,2}$ are real, the condition of the regular dynamics is:
$$
\cases{
2\varepsilon +\alpha +\beta >\sqrt{4\varepsilon ^2+
(\alpha -\beta )^2},\cr
\ \cr
\displaystyle \varepsilon +\frac{\alpha +\beta }{2}+\sqrt{\varepsilon ^2+
\left(\frac{\alpha -\beta }{2}\right)^2}<1.\cr
}
$$
The first inequality is always realized when
$\varepsilon ,\alpha ,\beta >0$. Thus, second one defines
the restriction for the region with regular dynamics.

\begin{figure}[ht]
\centering
\includegraphics[scale=0.8] {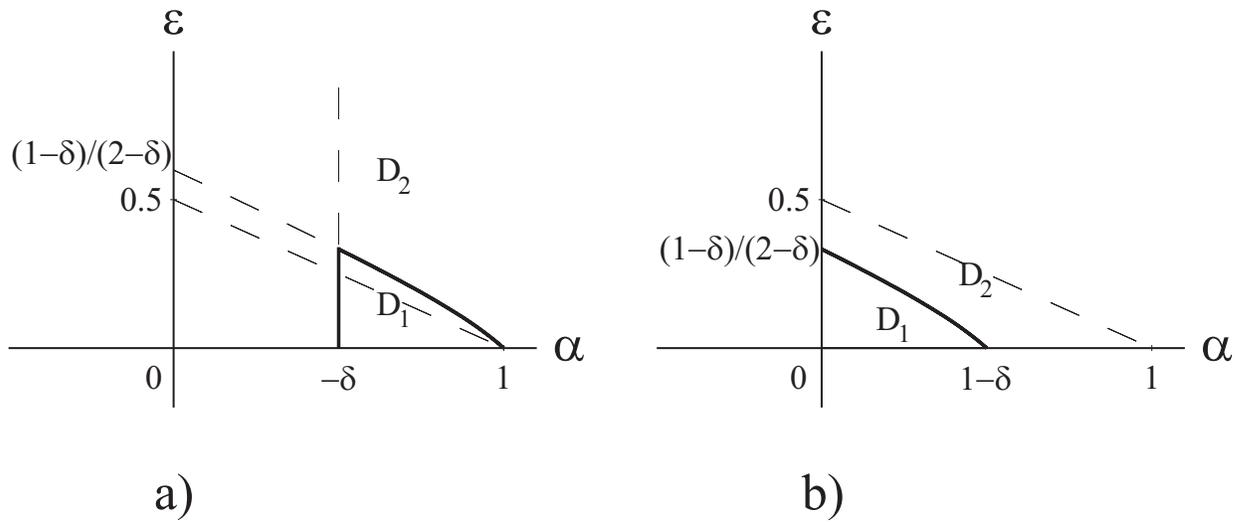}
\caption{Cross--sections by the planes $\delta={\rm const}$
of the parametric space of the model (\ref{g}), (\ref{period}) for
$-1<\delta<0$ (a) and $0<\delta<1$ (b).}
\label{D1D2per}
\end{figure}

In its turn, in the region of the parametric space where the Lyapunov
exponent $\lambda_s=\ln|\rho_s|>0$ the condition $\varepsilon
+(\alpha +\beta)/2+\sqrt{\varepsilon^2+ (\alpha -\beta )^2/4}>1$
is satisfied. Therefore, the boundary separating two regions with
regular and chaotic dynamics is a surface in the
three--dimensional parametric space $(\alpha, \beta,
\varepsilon)$, which is given by the relation:
\begin{equation}
\label{ten}
\varepsilon +\frac{\alpha +\beta }{2}+\sqrt{\varepsilon ^2+
\frac{(\alpha - \beta )^2}{4}}=1.
\end{equation}
To illustrate the obtained results we show in Fig.\ref{D1D2per}
cross--sections of this surface by the planes $\delta ={\rm
const}$, with $\delta=\beta -\alpha $ being the nonhomegeneity
parameter. Denote $D_1$ the region with regular dynamics and $D_2$
the region where $\lambda_s>0$. These two regions are separated by
the boundary, defined (implicitly) by Eq.(\ref{ten}). In
Fig.\ref{D1D2per} we also show by dashed line the corresponding
boundary between $D_1$ and  $D_2$ for the case of {\it
homogeneous}  $(\delta =0)$ system of coupled maps
(Fig.\ref{D1D2hom}).

As it follows from the Fig.\ref{D1D2per}, for $\delta >1$ for all values
of $\varepsilon$, $\alpha$ the model demonstrates the chaotic
dynamics. For $\delta <1$ the region $D_1$ arises where the
system has regular dynamics. This region where $0<\delta <1$
is, in fact, a subregion of the corresponding region $D_1$ of the
regular dynamics of the homogeneous model.

For $-1<\delta <0$ there exist such $\varepsilon$ and $\alpha$
that the homogeneous model exhibits the chaotic dynamics,
while the inhomogeneous still has the regular one. Note that the
restriction $\alpha >-\delta $ corresponds to the condition $\beta >0$.
For  $\delta <-1$ the region $D_1$ vanishes and for all $\varepsilon$,
$\alpha$ the model (\ref{period}), (\ref{g}) demonstrates only the
chaotic dynamics. As it also follows from Fig.\ref{D1D2per}, two
qualitatively different regimes make take place: $\alpha $ belongs
to the region of the regular behavior, $\beta$ to the region of chaos
with $\lambda_s>0$, with the global dynamics being regular. Similarly
one can observe just the opposite case of the chaotic dynamics
of the periodically inhomogeneous model. Realisation of the
particular regime depends on the value of the diffusion constant
$\varepsilon$.

\subsection{Annular model with a single defect}

Let us now turn to the case of spatially inhomogeneous model of
diffusively coupled maps for the case of a single ``defect''. In this
case among $N$ element of the system only one element has the
parameter $\beta$ (see (\ref{g})), while the other $(N-1)$ elements
have parameter  $\alpha$. Assume for simplicity that this element is
located at $n=1$. Then we can write for this model:
\begin{equation}
\label{defect}
\cases{
x_n(k+1)=G(x_n(k),\alpha ,\gamma )+
\varepsilon \Bigl(x_{n-1}(k)-2x_n(k)+
x_{n+1}(k)\Bigr), &$n=2,3,\ldots ,N$,\cr
\ \cr
x_1(k+1)=G(x_1(k),\beta,\gamma)+
\varepsilon \Bigl(x_N(k)-2x_1(k)+x_2(k)\Bigr),\cr
}
\end{equation}
where $ x_{n+N}\equiv x_n(k), n=1,2,\ldots ,N$. Let the function
$G(x,\alpha ,\gamma )$ is still given by (\ref{g}). Calculate
now the differential of the corresponding map for the
system (\ref{defect}), (\ref{g}):
\begin{equation}
\label{twelve}
D_f{\equiv }Q_N{=}\left(
\begin{array}{ccccc}
1-2\varepsilon -2\beta & \varepsilon & 0 & {\ldots }& \varepsilon \\
\varepsilon & 1-2\varepsilon -2\alpha & \varepsilon & {\ldots }& 0\\
0 & \varepsilon & 1-2\varepsilon -2\alpha & {\ldots }& 0\\
\vdots & \vdots & \vdots & {\ddots } & \vdots\\
\varepsilon & 0 & 0 & {\ldots }& 1-2\varepsilon -2\alpha \\
\end{array}
\right)\ .
\end{equation}
If we apply the same approach as previously, to find the Lyapunov
exponents of the model (\ref{defect}), (\ref{g}), we should have to
get the roots of the polynomial of the $2N+2$--th order. This can be
done only numerically. Instead we will use somewhat different
approach, which significantly facilitates the analysis.

Applying the estimate of the eigenvalues of the matrix (\ref{twelve})
according to the Gershgorin's theorem \cite{Horn}, we may evaluate
the location of the region of the regular dynamics. By this theorem, all
the eigenvalues of the matrix $A=\{ a_{ij}\} _{n\times n}$ which
belongs to the union of the discs on the complex plane:
$$
\rho _s\in\bigcup\limits_{i=1}^n
\Bigl\{ z\in {\bf C}:|z-a_{ij}|\le R'_i\Bigr\}\ ,
$$
where $s=1,2, \ldots ,N$  and  $R'_i$ is either a row almost--norm
of the matrix $A$, $R'_i=\sum_{j=1\atop j\ne i}^n |a_{ij}|$,
or a column almost--norm of the matrix $A$,
$R'_j=\sum_{i=1\atop i\ne j}^n |a_{ij}|$. Since the differential of the
map (\ref{twelve}) is a real symmetric matrix then, (i) the row and
column almost--norms are equal, and (ii) the eigenvalues of the matrix
${\rm D}_f$ are real. Therefore, the Gershgorin's discs transform
on the real axis as:
$$
\rho _s\in\bigcup\limits_{i=1}^n
\Bigl\{ z\in {\bf R}:|z-a_{ij}|\le R'_i\Bigr\}, \ \ \ \ s=1,2, \ldots ,N\ .
$$
Because $\varepsilon >0$, from (\ref{twelve}) we get:
$$
R'_i=\sum_{j=1\atop j\ne i}^n |a_{ij}|=|\varepsilon |+|\varepsilon |=
2\varepsilon\ .
$$

Moreover, since $R'_i$ does not depend on $i$ and in the differential
of map ${\rm D}_f$ (\ref{twelve}) there exist only two different diagonal
elements, the eigenvalues of ${\rm D}_f$ belong to the union of the
two intervals:
$$
\Bigl\{ z\in {\bf R}:|1-2\varepsilon -2\alpha -
z|\le 2\varepsilon \Bigr\} \bigcup \Bigl\{ z\in {\bf R}:|1-
2\varepsilon -2\beta -z|\le 2\varepsilon \Bigr\}.
$$
Then, taking into account that $\alpha$, $\beta$, $\varepsilon>0$,
we obtain the {\it upper} bound for the modules of the eigenvalues
of ${\rm D}_f$:
$$
|\rho _s|\le \cases{
\beta +2\varepsilon, &$\alpha <\beta $;\cr
\ &$\ $\cr
\alpha +2\varepsilon, &$\alpha >\beta $.\cr
}
$$
Thus, the region $\tilde D_1$ in the parametric space
$(\alpha,\beta,\varepsilon)$, which satisfies the conditions
\begin{equation}
\label{thirteen}
\tilde D_1:\qquad \cases{
\beta +2\varepsilon <1, & if $\alpha <\beta $;\cr
\ &$\ $\cr
\alpha +2\varepsilon <1, & if $\alpha >\beta $;\cr
}
\end{equation}
corresponds to the regular dynamics of the inhomogeneous model
(\ref{defect}), (\ref{g}), while the region
\begin{equation}
\label{fourteen}
\tilde D_2:\qquad \cases{
\beta +2\varepsilon >1, & if $\alpha <\beta $;\cr
&$\ $\cr
\alpha +2\varepsilon >1, & if $\alpha >\beta $;\cr
}
\end{equation}
corresponds to the positive $\lambda_s$.

\begin{figure}[ht]
\centering
\includegraphics[scale=0.7] {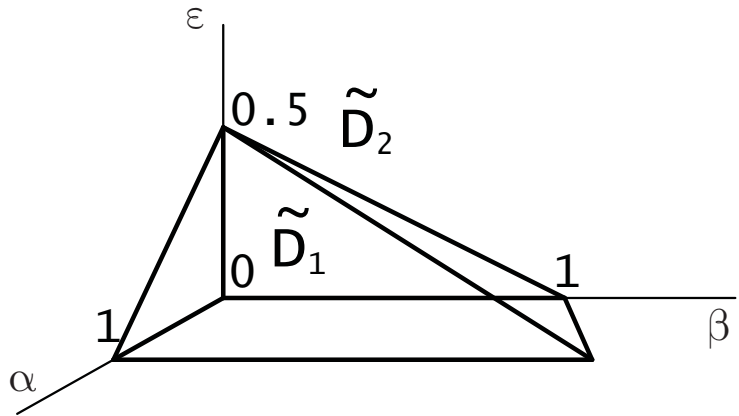}
\caption{The estimate of the region with the regular dynamics for
the inhomogeneous chains with two types of elements in the
parametric space $(\alpha, \beta,\varepsilon)$.}
\label{tildeD1D2in3D}
\end{figure}

The regions $\tilde D_1$ and $\tilde D_2$ are shown in
Fig.\ref{tildeD1D2in3D}. The region $\tilde D_1$ is the
{\it lower} bound of the region with the regular dynamics,
because it is obtained via the {\it upper} bound of the
eigenvalues of the map differential. This means that any
point of $\tilde D_1$ would correspond to the regular
dynamics, while any point of $\tilde D_2$ would correspond
either the regular dynamics, or (if the motion is finite) the chaotic one.

It is significant that the obtained estimate is applicable
not only for the model with a single defect, but for a {\it wide
class} of annular chains of  maps with diffusive coupling
(\ref{main}), which have only two types of elements
$\Big($with different parameters $\alpha _i $ in the function
$G(x_i, \alpha _i,\gamma)\Big)$. This follows from the fact
that the particular number of elements of each type and
their mutual location is not important. Indeed, almost--norms
of the differentials of the maps of all such systems are
the same and the centres of the two possible intervals, which contain
according to the Gershgorin's theorem, the eigenvalues of
${\rm D}_f$, do coincide.

In the case when the two intervals, which we use to estimate
$\rho_s$, do not intersect, there is, according to the Gershgorin's
theorem, ``clusterization'' of the eigenvalues of the differential of the
map. Namely, $q$ eigenvalues $\rho_s$, $s=1, 2,\ldots ,q$, will
belong to one interval, and $(N-q)$ eigenvalues $\rho _s$,
$s=q+1, q+2,\ldots,N$ will belong to the other interval. Here
$q$ is the number of elements with parameter $\alpha $, and,
respectively, $(N-q)$ is the number of elements with parameter
$\beta $. Unfortunately, this does not help to improve the estimate
of the region with the regular dynamics.

\begin{figure}[ht]
\centering
\includegraphics[scale=0.8] {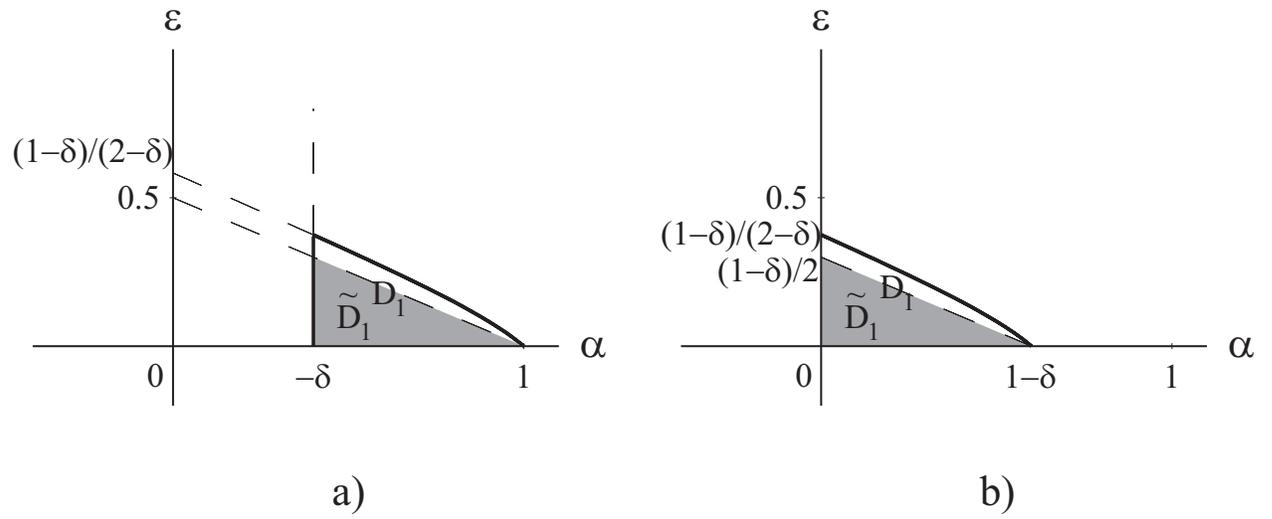}
\caption{The estimate and the exact result for the region with the
regular dynamics for the model with periodic nonhomegeneity for
$-1<\delta<0$ (a) and $0<\delta<1$ (b).}
\label{estimexper}
\end{figure}

Let us compare the estimate (\ref{thirteen}) with the exact result, that
has been obtained above for the model with the periodic
nonhomegeneity. Obviously, the latter model belongs to the
class of models for which the this estimate holds true. To illustrate
this we show the cross--sections $\delta ={\rm const}$ of the region
$D_1$, which corresponds to the regular dynamics of (\ref{period}),
(\ref{g}), and of its estimate, the region $\tilde D_1$ (see
Fig.\ref{estimexper}). From Fig.\ref{estimexper} it follows that the
estimate (\ref{thirteen}) of the region with the
regular dynamics is rather adequate, however it does not show that
for $-1<\delta <0$ the region of the regular dynamics of (\ref{period}),
(\ref{g}) includes the region of the regular dynamics of the
spatially homogeneous system.

\begin{figure}[ht]
\centering
\includegraphics[scale=0.8] {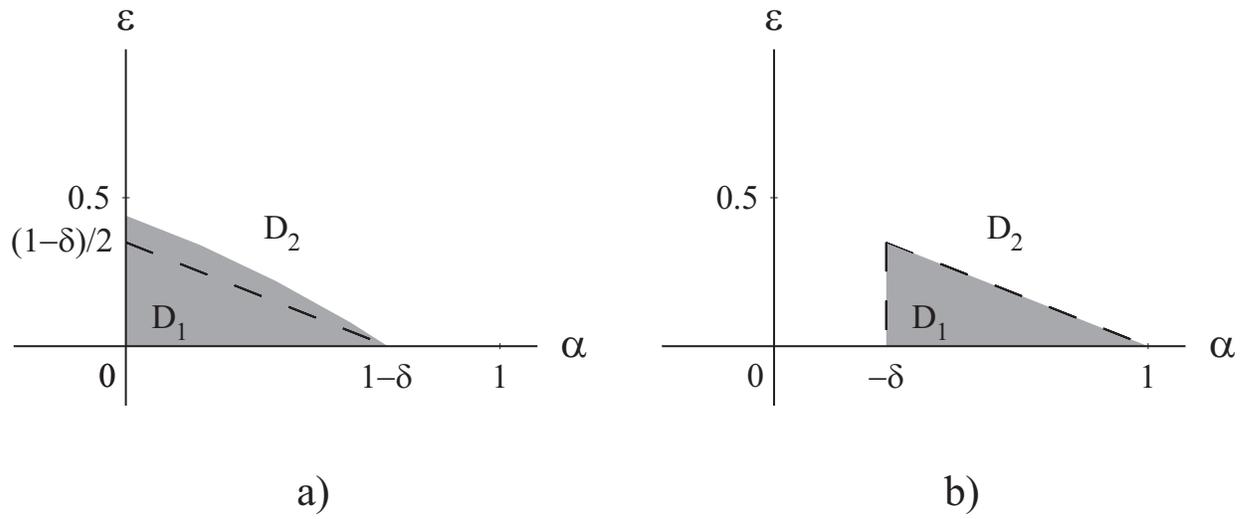}
\caption{Cross--sections by the planes $\delta={\rm const}$ of the
parametric space for the chain of $N=20$ elements with a single
defect for $\delta=-0.3$ (a) and  $\delta=0.3$ (b).}
\label{D1D2singl}
\end{figure}

Turn now to the map (\ref{defect}), (\ref{g}), which refer to the case
of a single defect and find the eigenvalues of the differential of map
(\ref{twelve}) numerically. One can write for the characteristic
polynomial:
$$
{\rm det}(Q_N-\rho I_N)=\left|
\begin{array}{ccccc}
2z\varepsilon +2(\alpha -\beta ) & \varepsilon & 0 & \ldots & \varepsilon \\
\varepsilon & 2z\varepsilon & \varepsilon & \ldots & 0\\
0 & \varepsilon & 2z\varepsilon & \ldots & 0\\
\vdots & \vdots & \vdots & \ddots & \vdots\\
\varepsilon & 0 & 0 & \ldots & 2z\varepsilon
\end{array}
\right|,
$$
where  $2z\varepsilon =1-2\varepsilon -2\alpha -\rho $.
Expanding ${\rm det}(Q_N-\rho I_N)$ in the first line we arrive
at
\begin{equation}
\label{fifteen}
{\rm det}(Q_N-\rho I_N)=[2z\varepsilon +2(\alpha -\beta )]B_{N-1}-
2\varepsilon ^2B_{N-2}-2(-1)^N\varepsilon ^N,
\end{equation}
with
$$
B_N=\left|
\begin{array}{ccccc}
2z\varepsilon & \varepsilon & 0 & \ldots & 0\\
\varepsilon & 2z\varepsilon &\varepsilon & \ldots & 0\\
0 & \varepsilon & 2z\varepsilon & \ldots & 0\\
\vdots & \vdots & \vdots & \ddots & \vdots\\
0 & 0 & 0 & \ldots & 2z\varepsilon
\end{array}
\right|.
$$
As previously, we use the first line factorization of $B_N$ to
obtain the recurrent relation, $B_N=2z\varepsilon B_{N-1}-
\varepsilon^2B_{N-2}$. Using the initial conditions,
$B_1=2z\varepsilon$, $B_2=\varepsilon ^2(4z^2-1)$,
we obtain:
\begin{equation}
\label{sixteen}
B_N = \varepsilon ^NU_N(z),
\end{equation}
where $U_N(z)$  is the second order Chebyshev polynomial.
Taking into account Eq.(\ref{sixteen}) and representation of
$U_N(z)$,
$$
U_N(z)=\frac{(z+\sqrt{z^2-1})^{N+1}-(z-\sqrt{z^2-1})^{N+1}}
{2\sqrt{z^2-1}}=
\frac{\left(\displaystyle \frac{1}{t}\right)^{N+1}-
t^{N+1}}{\displaystyle\frac{1}{t}-t}\ ,
$$
where $t=z-\sqrt{z^2-1}$, we recast Eq.(\ref{fifteen}) into the
following form:
$$
{\rm det}(Q_N-\rho I_N)=
2\varepsilon ^N\left[ \frac{t^{2N+1}+1}{2t^N}+\left( \frac{\alpha -\beta }
{\varepsilon }\right) \frac{1-t^{2N}}{(1-t^2)t^{N-1}}-(-1)^N\right] .
$$
Thus, the characteristic equation for the differential of the map
(\ref{twelve}) reads
$$
\displaystyle
t^{2N+2}+ 2\left(\displaystyle \frac{ \alpha - \beta }
{\varepsilon }\right) t^{2N+1} - t^{2N} -
2(-1)^Nt^{N+2} + 2(-1)^Nt^N + t^2 - 2\left(\displaystyle
\frac{\alpha - \beta}{\varepsilon}\right) t - 1=0.
$$
Finding numerically the roots of this equation,
we obtain $\rho_s$. Comparing $\rho _s$ with unity
for different sets of $\alpha$, $\beta$, $\varepsilon$, for the
annular chain of diffusively coupled maps with a single
defect we find the regions $D_1$ and  $D_2$  with qualitatively
different dynamics. In Fig.\ref{D1D2singl} we show cross--sections
of these regions by the planes with the fixed values of
$\delta={\rm const}$, where $\delta=\beta-\alpha$. The estimate
(\ref{thirteen}) of the region with the regular dynamics is marked by
the dashed line. As in the case of the periodic nonhomegeneity, the
actual region for positive $\delta$ is larger than that predicted by the
estimate. However for the negative $\delta$ the estimate is in a
perfect agreement with the numerical results as well as with the
results corresponding to the homogeneous chain. Thus, if one of the
elements of the chain had the parameter smaller than that of the
others, there would be no impact on the global dynamics of the
system. On the other hand, if the parameter which refers to the defect
had larger value than that of the other elements, this would diminish
the region of the regular dynamics as compared to the case of the
uniform system.

\section{Conclusions}

We investigated the dynamics of diffusively coupled one--dimensional
maps, which model one--dimensional excitable medium with
space nonhomegeneity. We considered different types of the
nonhomogeneities. For all systems, with the spatial nonhomegeneity,
which may be described in terms of two types of elements with
different values of the control parameter, we found {\it analytically}
a lower bound of the region in the parametric space where the
ensemble exhibits the regular dynamics. This estimate has been
obtained by localizing the eigenvalues of the differential of the map.
For two different model of this class we also obtained exact analytical
and numerical results which completely characterize the dynamics of
the system.

We observed that all three approaches (the approximate analytical
analysis of the chain with two types of elements (\ref{thirteen}),
the exact solution for the spatially periodic defects (\S2.2) and
numerical study (\S2.3) of the model with a single defect) show the
possibility of two qualitatively different types of the ensemble
behavior. In the first case the system demonstrates a regular global
behavior when one value of its parameter belongs to the region
with the regular dynamics and the other one belongs to the region
with the unstable dynamics. Vice versa, in the second case the global
dynamics occurs to be unstable, and if the motion is finite --- chaotic.
The particular type of the chain behavior depends on the diffusive
parameter $\varepsilon $. Obviously, if $\varepsilon $ is large enough,
the global dynamics would not be regular.

We wish to stress that according to our knowledge, the present
study is the first attempt of the {\it analytical} description
of the dynamics of inhomogeneous systems of coupled maps.
These systems, being of significant importance from the point
of view of applications, may exhibit a rich physical behavior,
giving rise to a lot of new and interesting problems. One can mention
such problems as properties of the wave travelling along the chain
and influence of the defects on these motions, dynamics of the
ensemble with time--dependent parameters, controllability
conditions, synchronization and many others (see, e.g.,
\cite{Parsin,Waxwa,Jopima,Kait,Bambiz,Mits}).
These problems, not discussed here will be addressed
elsewhere \cite{Lorybud,Lorybsel}.

\baselineskip=20pt

\newpage

\newpage

\newpage

\newpage

\newpage

\newpage

\end{document}